\documentclass[superscriptaddress,aps,prl,twocolumn]{revtex4}
\usepackage[utf8]{inputenc}
\usepackage[T1]{fontenc}
\usepackage{bm}
\usepackage{ulem}
\usepackage{epsfig}
\usepackage{graphicx}
\usepackage{amssymb,amsmath,amsbsy,amsgen,amsfonts}    
\usepackage{dcolumn}
\usepackage{amsthm}
\usepackage{mathrsfs}
\usepackage{latexsym}
\usepackage{array}
\usepackage{color}
\usepackage{amstext}
\allowdisplaybreaks[1]
\usepackage{txfonts}
\usepackage{pifont}

\usepackage{epstopdf} %for texshop on mac

\newcommand{\ket}[1]{\left\vert{#1}\right\rangle}
\newcommand{\ketbra}[2]{|#1\rangle \langle#2|}

\newcommand{\be}{\begin{equation}}
\newcommand{\ee}{\end{equation}}
\newcommand{\ba}{\begin{array}}
\newcommand{\ea}{\end{array}}
\newcommand{\bqa}{\begin{eqnarray}}
\newcommand{\eqa}{\end{eqnarray}}

\DeclareSymbolFont{symbols}{OMS}{cmsy}{m}{n}

\begin{document}

\title{Experimental investigation of Markovian and non-Markovian channel addition}

\author{S. A. Uriri}
\affiliation{School of Chemistry and Physics, University of KwaZulu-Natal, Durban 4001, South Africa}
\author{F. Wudarski}
\email{fawudarski@gmail.com}
\affiliation{Quantum Artificial Intelligence Lab. (QuAIL), Exploration Technology Directorate, NASA Ames Research Center, Moffett Field, CA 94035, USA} 
\affiliation{USRA Research Institute for Advanced Computer Science (RIACS), Mountain View, CA 94035, USA}
\affiliation{Institute of Physics, Faculty of Physics, Astronomy and Informatics, Nicolaus Copernicus University, Grudzi\k{a}dzka 5/7, 87–100 Toru\'n, Poland}
\author{I. Sinayskiy}
\affiliation{School of Chemistry and Physics, University of KwaZulu-Natal, Durban 4001, South Africa}
\author{F. Petruccione}
\affiliation{School of Chemistry and Physics, University of KwaZulu-Natal, Durban 4001, South Africa}
\affiliation{National Institute for Theoretical Physics, KwaZulu-Natal, South Africa}
\author{M. S. Tame}
\email{markstame@gmail.com}
\affiliation{Department of Physics, Stellenbosch University, Matieland 7602, South Africa}

\date{\today}

\begin{abstract}
The study of memory effects in quantum channels helps in developing characterization methods for open quantum systems and strategies for quantum error correction. Two main sets of channels exist, corresponding to system dynamics with no memory (Markovian) and with memory (non-Markovian). Interestingly, these sets have a non-convex geometry, allowing one to form a channel with memory from the addition of memoryless channels and vice-versa. Here, we experimentally investigate this non-convexity in a photonic setup by subjecting a single qubit to a convex combination of Markovian and non-Markovian channels. We use both divisibility and distinguishability as criteria for the classification of memory effects, with associated measures. Our results highlight some practical considerations that may need to be taken into account when using memory criteria to study system dynamics given by the addition of Markovian and non-Markovian channels in experiments.
\end{abstract}

%\pacs{}

\maketitle

%%%%%%%%%%%%%%%%%%%%%%%%%%%%
%%%%%%%%%%%%%%%%%%%%%%%%%%%%
%%%%%%%%%%%%%%%%%%%%%%%%%%%%
%%%%%%%%%%%%%%%%%%%%%%%%%%%%
\section{I. Introduction} 

Quantum systems interact with their environment in a number of ways, leading to adverse effects such as decoherence~\cite{Schloss05} and noise~\cite{Clerk10}. In quantum information processing -- for instance in quantum communication~\cite{Gis02} and quantum computing~\cite{Ladd10} -- this may result in a loss of quality of states and their correlations in a system, which can have a detrimental effect on the performance of a given task. The study of the evolution of quantum systems open to their environment provides a deeper understanding of these effects~\cite{Breuer02,Lidar19}, where the dynamics can be described with the use of quantum channel theory~\cite{Caruso14}. Here, there are two main sets of channels: those with no memory effects, or back flow of information from the environment to the system, known as Markovian, and those with memory effects, where information can flow back into the system from the environment, known as non-Markovian. Contrary to classical stochastic processes, quantum Markovianity lacks a unique definition, and various not necessarily equivalent formulations coexist \cite{Vacchini11}. The two leading avenues to characterize Markovianity are based on: (i) a quantum maps and master equation approach~\cite{Rivas10,Breuer09,Lu10,Rajagopal10,Luo12,Bylicka14,Lorenzo13}, and (ii) modelling of the full system and environment dynamics \cite{Li18,LoGullo14,Pollock18}. In recent years, both approaches have attracted considerable attention from the scientific community~\cite{Rivas14,Vega17,Breuer16,Aolita15,LoFranco13,Tang12,Bernardes15,Xu13,Man15a,Man15b,Brito15}, driven by a desire for gaining a deeper theoretical understanding of quantum memory effects and experimental advances in the quantum control of various physical systems, such as those using photons~\cite{OBrien05,Flamini19}, atoms~\cite{Blatt12,Muller12} and superconducting settings~\cite{Clarke08,Wendin17}. 

While the states of quantum systems form a convex set, where any state can be formed from the addition of other states~\cite{Nielsen00}, when using the quantum maps and master equation approach we have that Markovian and non-Markovian channels describing the dynamics of those quantum systems fail to form such a set~\cite{Wolf08,Megier17}. In the past, several works have studied this non-convex geometry and introduced interesting examples where the addition of Markovian channels leads to a non-Markovian channel~\cite{Wolf08,Wudarski17,Shrikant18,Breuer17,Wudarski15} and vice versa~\cite{Wudarski16}. However, despite many experiments realising instances of quantum channels~\cite{Bongioanni10,Piani11,Fisher12,Lu15,Marques15,Wang15,Liu18,McCutcheon18,Liu18} and specifically non-Markovian channels~\cite{Liu11,Chiuri12,Cialdi17,Morris19,Wu19}, so far there has been no experimental investigation of the interesting phenomenon that adding Markovian channels can give rise to a non-Markovian channel, or the other way around.

\begin{figure*}[t]
\includegraphics[width=17.5cm]{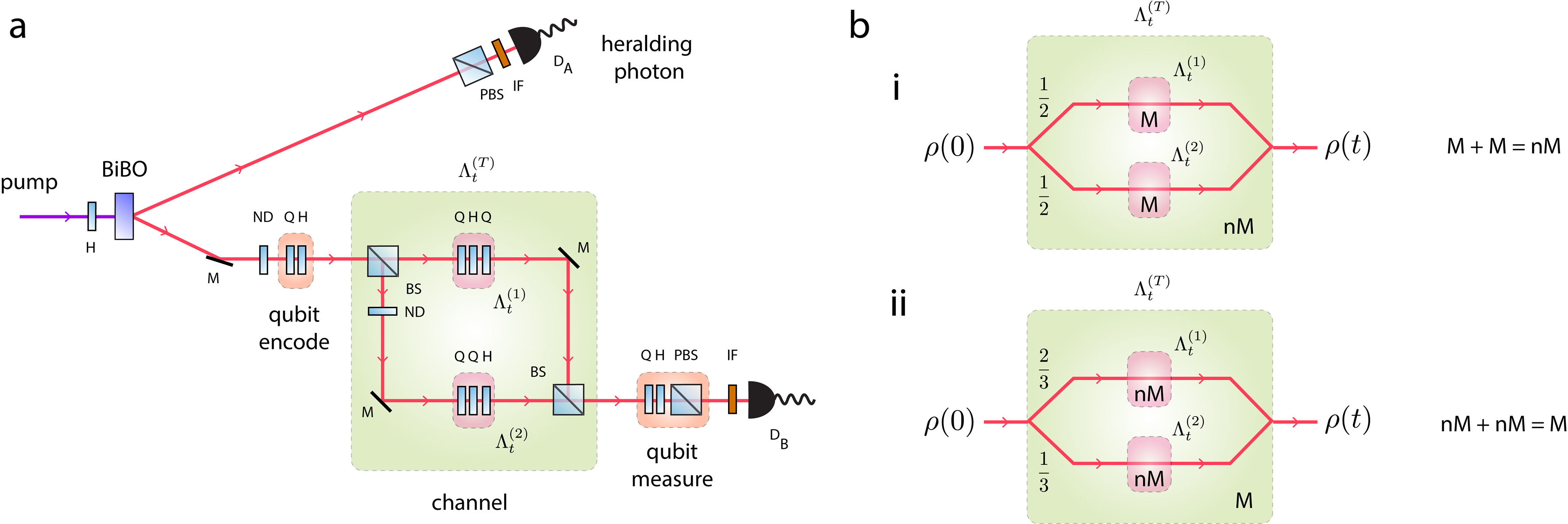}
\caption{Experimental setup for the implementation of Markovian and non-Markovian channel addition. {\bf (a)} Linear optics setup, where a nonlinear BiBO crystal is pumped with a CW laser at 405 nm, producing pairs of photons at 810 nm via spontaneous parametric down-conversion. One photon is detected at detector {\sf D}$_{\sf A}$ and heralds the presence of a single photon in the other arm, which is used to encode a qubit in the polarization degree of freedom. The input qubit is split probabilistically into two paths by a beamsplitter, with different polarization operations applied for realizing different channels. The individual channels ($\Lambda_t^{(1)}$ and $\Lambda_t^{(2)}$) are then added together by recombining the paths using a second beamsplitter (which incurs 50\% loss overall) to give rise to a total channel $\Lambda_t^{(T)}$. The state of the output qubit is measured in the polarization basis. The rate of input qubits is reduced using the first neutral density filter to ensure that at most one qubit is in the split region at any one time. Here {\sf H} is a half-wave plate, {\sf Q} is a quarter-wave plate, {\sf PBS} is a polarizing beam splitter, {\sf BS} is a beamsplitter, {\sf ND} is a neutral density filter, {\sf IF} is an interference filter ($\lambda_0$ = 810 nm and  $\Delta\lambda$ = 10 nm), and {\sf D}$_{\sf A}$ and {\sf D}$_{\sf B}$ are avalanche single-photon detectors. {\bf (b)} Pictorial representation of the Markovian and non-Markovian channels obtained by addition using the experimental setup. Panel (i) depicts the equal addition of two Markovian channels, giving rise to a non-Markovian channel. Panel (ii) depicts the unequal addition of two non-Markovian channels, giving rise to a Markovian channel. The unequal addition is achieved in the setup by controlling the probability of splitting the two paths using a second {\sf ND} filter.}
\label{fig1} 
\end{figure*}

In this work, we investigate experimentally the mixing of Markovian and non-Markovian channels in a photonic setup using the quantum maps and master equation approach. By encoding a single qubit into a single photon in the polarization degree of freedom, we implement the addition of two Markovian channels using linear optics and study the extent to which these channels are Markovian and the resulting channel is non-Markovian. We also implement the reverse scenario of the convex combination of two non-Markovian channels and study the resulting channel's Markovian nature. We use both divisibility~\cite{Rivas10} and distinguishability~\cite{Breuer09} as criteria for the classification of memory effects, along with their associated measures. We find that for the examples realized, the addition of channels leads to important practical considerations that must be taken into account when using memory criteria and their measures for specific types of channels in experiments. The results may help in the theoretical development of more robust criteria and measures for the assessment of Markovian and non-Markovian effects in experimental quantum systems where both sets of channels are present.

The paper is structured as follows. In Section II, we introduce the experimental setup and show how it can be used to implement the addition of Markovian and non-Markovian channels based on the theory. In Section III, we discuss the results of the experiment. In Section IV we summarize our findings. 

%%%%%%%%%%%%%%%%%%%%%%%%%%%%
%%%%%%%%%%%%%%%%%%%%%%%%%%%%
%%%%%%%%%%%%%%%%%%%%%%%%%%%%
%%%%%%%%%%%%%%%%%%%%%%%%%%%%
\section{II. Experimental setup} 

In Fig.~\ref{fig1}~(a) we show the experimental setup used to implement the addition of Markovian and non-Markovian channels. Here, single photons are generated using a heralded type-1 spontaneous parametric down-conversion (SPDC) source~\cite{Burnham70,Hong1986}. A continuous wave laser with a wavelength of 405 nm (Coherent OBIS 405 nm) has its polarization set to vertical by a half-wave plate (HWP) and is used to pump a non-linear BiBO crystal. The SPDC process produces two `twin' photons (idler and signal) polarized horizontally and at a lower frequency (wavelength 810 nm), with one photon produced in arm A (top path) and the other in arm B (bottom path). The optical axis of the BiBO crystal is cut such that the two photons emerge at $\pm3^0$ from the initial pump direction. Each arm has an interference filter ($\lambda_0$ = 810 nm and  $\Delta\lambda$ = 10 nm) placed before single-photon detectors (Excelitas SPCM-AQRH-15), denoted as ${\sf D_A}$ and ${\sf D_B}$. The filters enable the photons to be spectrally selected from the SPDC emission and well defined in terms of their spectrum. A single photon in arm A with horizontal polarization is transmitted by the polarizing beamsplitter (PBS) and its detection is used to `herald' the presence of the other single photon of the SPDC pair in arm B. This heralded photon has a single qubit encoded into its polarization degree of freedom using a quarter-wave plate (QWP) and HWP, where the computational basis is represented by horizontal and vertical polarization, {\it i.e.} $\{ \ket{H},\ket{V}\}$. The remainder of the setup implements the addition of Markovian and non-Markovian channels, which we now describe.

In the experiment we aim to implement two separate cases for the addition of Markovian and non-Markovian channels: (i) Two Markovian channels added to produce a non-Markovian channel, which we denote as `${\sf M+M=nM}$', and (ii) Two non-Markovian channels added to produce a Markovian channel, denoted as `${\sf nM+nM=M}$'. These two cases are shown in Fig.~\ref{fig1}~(b). In order to implement them in the setup, we split the encoded input qubit into two paths using a beamsplitter (BS), where each path has a particular type of channel realized using QWPs and HWPs. The paths are then combined at a second beamsplitter (which incurs a 50\% loss) and the output qubit is measured. It is important to note that while half of the photons are lost at the second BS, this does not affect the implementation of the channels, as they are performed in the polarization basis. The loss is unbiased to which channel the photon went through and thus it simply reduces the rate at which the addition of channels is implemented. Furthermore, the paths that are split and recombined by the BSs are not interferometrically aligned temporally (as in a Mach-Zehnder interferometer for instance). This enables the channels implemented on either path to be simply added incoherently through spatial alignment as a linear summation, as required in theory. Thus, there is no superposition of the photon (or qubit) in the split region -- the output state is a statistical mixture of the input state having one of two channels applied. This is aligned with a collision model realizing the summation of dynamical maps~\cite{Filippov17}. Finally, we use a neutral density filter to reduce the rate of single qubits entering the split region in order to ensure at most one qubit is present in either channel at any one time.
\begin{figure}[t]
\includegraphics[width=8.6cm]{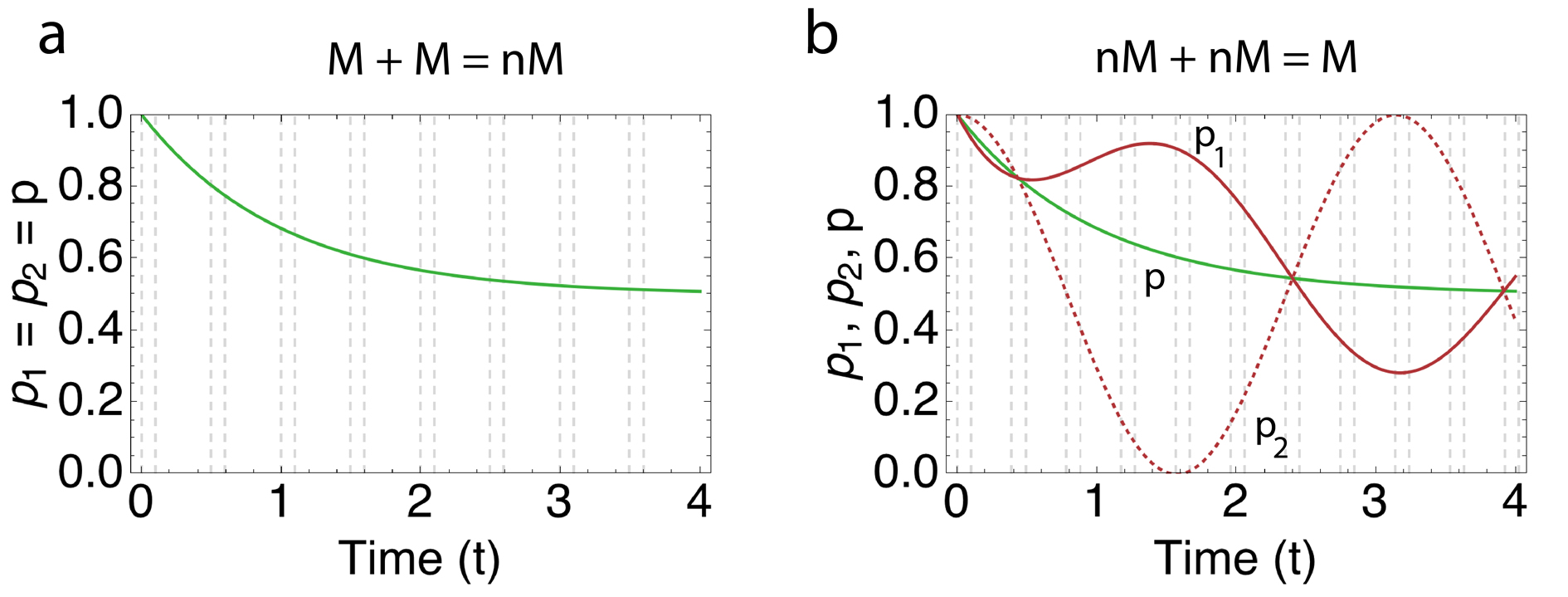}
\caption{Markovian and non-Markovian channel addition. {\bf (a)} Theoretical probability, $p(t)$, for the application of noise operators in the individual (Markovian) and total (non-Markovian) channels for ${\sf M+M=nM}$. {\bf (b)} Theoretical probability, $p_i(t)$, for the application of noise operators in the individual (non-Markovian) and total (Markovian) channels for ${\sf nM+nM=M}$. The dotted lines correspond to times at which the channels are implemented in the experiment.}
\label{fig2} 
\end{figure}

\subsection{A. Markovian channel addition}

The first case implemented is the addition of two Markovian channels to make a non-Markovian channel, as shown in Fig.~\ref{fig1}~(b)~(i). Here, the individual Markovian channels on the top (1) and bottom (2) paths correspond to dynamics that a quantum system is subjected to for a given time duration. In this sense, by realizing the channels in our setup we are effectively simulating the dynamics that a quantum system undergoes for different time durations. Consider an initial time $t=0$, where the state of the system is $\rho(0)$ and a later time $t$ where the state is given by $\rho(t)=\Lambda_t^{(i)} (\rho(0))$, and $\Lambda_t^{(i)}$ is a completely positive trace-preserving (CPTP) map that represents channel $i$, corresponding to some system dynamics that occur over a time period $t$. The two channels we consider here are given explicitly as
\bqa
\Lambda_t^{(1)}(\rho)&=& p(t)\rho +(1-p(t)) \sigma_x \rho \sigma_x, \label{ch11} \\
\Lambda_t^{(2)}(\rho)&=& p(t)\rho +(1-p(t)) \sigma_y \rho \sigma_y, \label{ch12}
\eqa
where $\rho=\rho(0)$ and $\sigma_i$ are the usual Pauli matrices. These channels represent `phase' damping along the $x$ and $y$ axes of the Bloch sphere, respectively~\cite{Nielsen00}. Here, the probability $p(t)=(1+ e^{-t})/2$ corresponds to a time evolving probability in the system dynamics, which is shown in Fig.~\ref{fig2}~(a). By adding the two channels as an equal linear summation, {\it i.e.} $\Lambda_t^{(T)}(\rho)=\frac{1}{2}(\Lambda_t^{(1)}(\rho)+\Lambda_t^{(2)}(\rho))$, one obtains the total channel
\be
\Lambda_t^{(T)}(\rho)= p(t)\rho +\frac{1}{2}(1-p(t))( \sigma_x \rho \sigma_x + \sigma_y \rho \sigma_y).
\label{ch1T}
\ee
The individual channels $\Lambda_t^{(1)}$ and $\Lambda_t^{(2)}$ are contained within the Markovian set of channels according to the completely positive (CP) divisibility criteria~\cite{Rivas14,Vega17,Li18,Wudarski17}, because for any pair of intermediate times $(s,t)$ during the system dynamics we can write 
\be
\Lambda_t^{(i)}(\rho)=V_{t,s}^{(i)}(\Lambda_s^{(i)}(\rho))~~, \qquad t \geq s,
\label{interm}
\ee
where $V_{t,s}^{(i)}$ is a CP operator describing the intermediate map from $s$ to $t$. On the other hand, the time evolution from zero to $s$ may induce correlations between the system and its environment such that $V_{t,s}^{(i)}$ is not a CP operator, even though $\Lambda_t^{(i)}$ is CP. In this case, the dynamics are not `divisible' in terms of complete positivity ($\Lambda_t^{(i)}\neq V_{t,s}^{(i)}\Lambda_s^{(i)}$, where $V_{t,s}^{(i)}$ is CP) and the channel $\Lambda_t^{(i)}$ belongs to the non-Markovian set of channels. This is the case for the above total channel $\Lambda_t^{(T)}$ for any pair of times~\cite{Wudarski17}. It is interesting to note that recent work~\cite{Milz19} has shown that non-Markovian temporal correlations may be able to hide in a divisible process and therefore Markovian and CP divisibility do not always coincide with each other in a stricter operational sense. However, as shown in Ref.~\cite{Milz19}, if the channel is CP divisible it can be seen as one that is Markovian on average.

In the experimental setup, the addition of the individual channels $\Lambda_t^{(1)}$ and $\Lambda_t^{(2)}$ to produce the total channel $\Lambda_t^{(T)}$ is achieved by using a pair of BSs -- the first splits the input state $\rho(0)$ probabilistically into two paths where the individual channels are applied and the second recombines the channels to give the output state $\rho(t)$ for some time duration $t$ corresponding to the system dynamics. We implement the channels for a fixed set of time durations, from $t=0$ to $t=3.6$ in coarse steps of 0.5 (with finer steps of 0.1), shown as dotted lines in Fig.~\ref{fig2}~(a). 

For a specific time duration $t$ there is a corresponding value of $p(t)$ that determines the probabilistic application of noise in the individual channels -- $\sigma_x$ in channel 1 and $\sigma_y$ in channel 2. We realize this noise by using automated wave plates in the optical paths. For channel 1, we use a QWP-HWP-QWP chain, where the angles of the QWPs are set to zero and the angle of the HWP is set to zero for the identity operation and  45$^\circ$ for $\sigma_x$. For channel 2, we use a QWP-QWP-HWP chain, where again the angles of the QWPs are set to zero and the angle of the HWP is set to zero for the identity and 45$^\circ$ for $\sigma_y$. For a given time duration $t$ of the system the value $p(t)$ is set and we probe the split region with an input qubit $\rho(0)$, measuring the output qubit $\rho(t)$ over many repetitions in order to build up statistics from the measurements. During the repetitions, the HWP angles on each path are modified probabilistically and independently, with the ratio of occurrence of the angles being zero or 45$^\circ$ given by the value of $p(t)$. Thus, the individual channels are realized in the paths and the total channel is given by their probabilistic addition, as required by Eq.~(\ref{ch1T}).

\subsection{B. Non-Markovian channel addition}

The second case implemented is the addition of two non-Markovian channels to make a Markovian channel, as shown in Fig.~\ref{fig1}~(b)~(ii). Here, the individual channels on the top (1) and bottom (2) paths are described by the CPTP maps
\bqa
\Lambda_t^{(1)}(\rho)&=& p_1(t)\rho +(1-p_1(t)) \sigma_x \rho \sigma_x, \label{ch21} \\
\Lambda_t^{(2)}(\rho)&=& p_2(t)\rho +(1-p_2(t)) \sigma_x \rho \sigma_x, \label{ch22}
\eqa
where the probabilities are $p_1= 3 [ (1+e^{-t})/2 -\cos^2 (t)/3 ]/2$ and $p_2=\cos^2 (t)$, which are shown as the red solid and red dotted lines in Fig.~\ref{fig2}~(b), respectively. For particular time pairs $(s,t)$, both these channels belong to the non-Markovian set of channels, in terms of CP divisibility. We give an explicit example of such a time pair in the next section. On the other hand, by adding the two channels as a weighted linear summation, given as $\Lambda_t^{(T)}(\rho)=\frac{2}{3}\Lambda_t^{(1)}(\rho)+\frac{1}{3}\Lambda_t^{(2)}(\rho)$, one obtains the total channel
\be
\Lambda_t^{(T)}(\rho)= p(t)\rho +(1-p(t)) \sigma_x \rho \sigma_x,
\ee
where $p(t)=(1+ e^{-t})/2$, shown in Fig.~\ref{fig2}~(b) as a green solid line. This channel represents phase damping along the $x$ axis of the Bloch sphere and is Markovian (semigroup~\cite{Gorini76,Lindblad76}) for any pair of times~\cite{Wudarski16}.

In the experimental setup, the addition of the individual channels $\Lambda_t^{(1)}$ and $\Lambda_t^{(2)}$ is again achieved by using a pair of BSs. However, in order to achieve the weighted addition we use a neutral density filter to control the transmission in path 2, with the ratio between the two paths set to be equal to $2:1$, as shown in Fig.~\ref{fig1}~(a). We implement the individual and total channels for a fixed set of time durations, from $t=0$ to $t=5\pi/4+0.1$, shown as dotted lines in Fig.~\ref{fig2}~(b). For a specific time duration $t$ of the dynamics there are corresponding values of $p_i(t)$ that determine the probabilistic application of noise in the individual channels (in Eqs.~(\ref{ch21}) and (\ref{ch22})). As before, we realize the noise by using waveplates in the optical paths. For both channels, a QWP-HWP-QWP chain is used, where the angles of the QWPs are set to zero and the angle of the HWP is set to zero for the identity operation and 45$^\circ$ for $\sigma_x$. When probing the total channel, the HWP angles on each path are modified probabilistically and independently, with the ratio of occurrence of the angles being zero or 45$^\circ$ for channel $i$ given by the value of $p_i(t)$.

%%%%%%%%%%%%%%%%%%%%%%%%%%%%
%%%%%%%%%%%%%%%%%%%%%%%%%%%%
%%%%%%%%%%%%%%%%%%%%%%%%%%%%
%%%%%%%%%%%%%%%%%%%%%%%%%%%%
\section{III. Results}

%%%%%%%%%%%%%%%%%%%%%%%%%%%%
%%%%%%%%%%%%%%%%%%%%%%%%%%%%
\subsection{A. Characterization method}

In order to determine whether the individual and total channels are Markovian or non-Markovian, we perform quantum process tomography~\cite{Chuang97} for a fixed set of time durations and obtain the corresponding $\chi$ matrices for the channels. With these matrices we are then able to check for divisibility via positivity of the Choi matrix~\cite{Jam72,Choi75} representing the map between two intermediate times and therefore the Markovian or non-Markovian nature of the channels~\cite{Wudarski17}. We now briefly describe this method before discussing the results. 
\begin{figure*}[t]
\includegraphics[width=17.5cm]{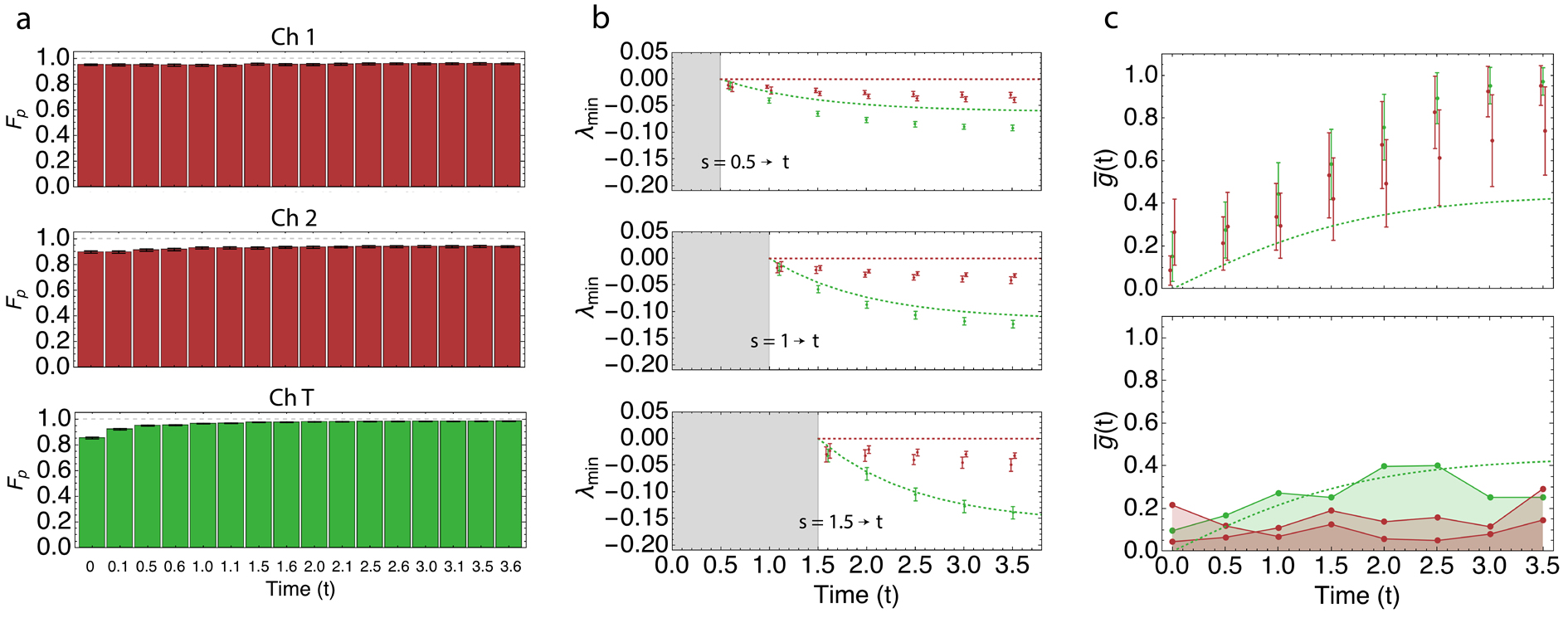}
\caption{Markovian channel addition: ${\sf M+M=nM}$. {\bf (a)} Process fidelities, $F_p$, for the $\chi$ matrices of the realized channels for the times corresponding to the dotted lines in Fig.~\ref{fig2}~(a). {\bf (b)} Minimum eigenvalues, $\lambda_{\rm min}$, of the Choi matrix for the intermediate maps from time $s$ to $t$, for $s=0.5$ (upper plot), $1$ (middle plot) and $1.5$ (lower plot). The dashed grey areas highlight that $t \geq s$ must be satisfied. {\bf (c)} Top plot shows the function $\bar{g}(t)$ whose area corresponds to the amount of non-Markovianity accumulated during the dynamics. The values and error bars are obtained by pairing up each of the 100 $\chi$ matrices from the experiment for time $t$ and the 100 $\chi$ matrices for $t+0.1$, giving $10,000$ Choi matrices for the intermediate map. The bottom plot shows $\bar{g}(t)$ calculated using the average $\chi$ matrices for times $t$ and $t+0.1$. In panels (b) and (c), the red (green) points correspond to the values from the experiment for the individual (total) channels. The red (green) dotted lines correspond to the ideal theoretical values for the individual (total) channels. The data points for the individual channels have been shifted slightly in time away from each other for clarity.}
\label{fig3} 
\end{figure*}

For a given channel we have that the input state $\rho(0)$ evolves to $\rho(t)=\Lambda_t(\rho(0))= \sum_{m,n} \chi_{mn} \hat{E}_m \rho(0) \hat{E}_n^\dag$, where $\{ \hat{E}_i\}$ are a fixed set of operators forming a complete basis for the Hilbert space, {\it i.e.} for a single qubit the set can be made from the Pauli matrices and the identity operator: $\{ \openone,\sigma_x,\sigma_y,\sigma_z \}$. The $\chi$ matrix then represents a complete description of the quantum channel for a fixed time $t$. We obtain the elements of the $\chi$ matrices by encoding the input qubit $\rho(0)$ as one of the probe states $\ket{H}$, $\ket{V}$, $\ket{+}=(\ket{H}+\ket{V})/\sqrt{2}$ and $\ket{+_y}=(\ket{H}+i\ket{V})/\sqrt{2}$ which is then sent through the individual channels (by blocking one path or the other) and the total channel (both paths combined)~\cite{Chuang97}. Each probe state is sent through a given channel many times and quantum state tomography~\cite{James01} is performed on the output state while the appropriate noise operations are applied. From the probe states the $\chi$ matrix is then obtained using a maximum likelihood reconstruction~\cite{Chuang97}.

With a knowledge of the $\chi$ matrices for two different time durations $s$ and $t$ for a given channel we then check the positivity of the intermediate map linking $s$ to $t$ (see Eq.~(\ref{interm})), {\it i.e.} $V_{t,s}$. To do this, we construct the transfer matrices $F(s)$ and $F(t)$ for the maps $\Lambda_{s}$ and $\Lambda_{t}$ from the $\chi$ matrices for $s$ and $t$, respectively. The transfer matrix approach is a useful technique that allows a density matrix to be represented as a stacked vector, $|| \rho \rangle$, with the evolution of the system written as $|| \rho(t) \rangle=F(t) || \rho(0) \rangle$. The elements of a transfer matrix $F(x)$ are given explicitly as
\be
F_{\alpha,\beta}(x)={\rm Tr} (G_\alpha^\dag \Lambda_x(G_\beta)),
\ee
where $\{G_\alpha \}$ is a set of orthonormal operators with respect to the Hilbert-Schmidt inner product, chosen here to be the unit matrices basis $\{G_1= \ketbra{H}{H},G_2=\ketbra{H}{V},G_3=\ketbra{V}{H},G_4=\ketbra{V}{V}\}$ for the single qubit case. With a knowledge of the $\chi$ matrix for a given time $x$ we can calculate $\Lambda_x(G_\beta)$ and obtain the transfer matrix $F(x)$. From the transfer matrices for two different times $s$ and $t$ we obtain the transfer matrix $F(t,s)=F(t)F(s)^{-1}$ for the intermediate map $V_{t,s}$. Using this we then form the Choi matrix, which for a given transfer matrix $F(x)$ for a qubit is written as
\be
W(x)=\frac{1}{2} \sum_{\alpha,\beta=1}^4 F_{\alpha,\beta}(x) G_\beta \otimes G_\alpha,
\ee
obtained simply by applying the dynamical map $\Lambda_x$ to one qubit of the maximally entangled state $\ket{\psi_+}=\frac{1}{\sqrt{2}}(\ket{H}\otimes \ket{H}+\ket{V}\otimes \ket{V})$, {\it i.e.} $W(x)= (\openone \otimes \Lambda_x)P_+$, with $P_+=\ketbra{\psi_+}{\psi_+}$. Due to the Choi-Jamio\l kowksi isomorphism~\cite{Jam72,Choi75} a dynamical map $\Lambda_x$ is CP if and only if the corresponding Choi matrix is positive. In other words, if any of the eigenvalues of the Choi matrix for an intermediate map between times $s$ and $t$ are negative then the dynamics are not divisible and the channel is non-Markovian, otherwise the channel is Markovian. We use this criteria in what follows for the two cases investigated.

%%%%%%%%%%%%%%%%%%%%%%%%%%%%
%%%%%%%%%%%%%%%%%%%%%%%%%%%%
\subsection{B. Markovian channel addition}

In this first case, we follow the experimental procedure outlined in the previous section in order to implement the addition of two Markovian channels to give a non-Markovian channel. As mentioned, the channels are realized for a set of fixed times and quantum process tomography is carried out to obtain the $\chi$ matrices. For the set of times marked in Fig.~\ref{fig2}~(a) we show the corresponding process fidelities for the channels in Fig.~\ref{fig3}~(a), where the process fidelity, $F_p={\rm Tr}(\sqrt{\sqrt{\chi}\chi_{id}\sqrt{\chi}})^2 / {\rm Tr}(\chi){\rm Tr}(\chi_{id})$~\cite{Jozsa94}, quantifies how close the experimental channel $\chi$ is to the ideal theoretical channel $\chi_{id}$. A value of $F_p=1$ corresponds to a perfect overlap, whereas $F_p=0$ corresponds to zero overlap between the experiment and theory. One can see that the experimental channels are of high quality for both the individual channels (Ch 1 and Ch 2) and total channel (Ch T), with the majority of process fidelity values $>0.90$. For each fixed time, the error in $F_p$ is obtained from 100 $\chi$ matrices generated from the measured counts with Poissonian fluctuations in the count statistics~\cite{James01}. 

The 100 $\chi$ matrices for a fixed time result in 10,000 pairs of $\chi$ matrices for each time pair $(s,t)$. The 10,000 Choi matrices for the intermediate maps are then calculated, as described in the previous subsection, and their eigenvalues are obtained. The lowest eigenvalue $\lambda_{\rm min}$ is plotted in Fig.~\ref{fig3}~(b) for time pairs starting with $s=0.5$, 1 and $1.5$, and $t>s$. One can see that for all time pairs $\lambda_{\rm min}$ for the total channel is always below zero, confirming that the channel is not divisible and therefore non-Markovian. Indeed, $\lambda_{\rm min}$ follows the theoretically expected behavior (green dotted line), with a better matching observed as $s$ increases. This may be attributed to $\lambda_{\rm min}$ becoming more negative as $s$ increases and therefore the values are less influenced by the non-ideal conditions in the experiment. 

On the other hand, for the individual channels $\lambda_{\rm min}$ is also negative for any time pair. Here, the data points of channels 1 and 2 are shifted slightly off-center from each other for a given time $t$ for clarity. The negative $\lambda_{\rm min}$ for the individual channels is unexpected as they should be Markovian ideally and therefore the Choi matrix should have no negative eigenvalues. The issue here is that in the theoretical ideal case $\lambda_{\rm min}=0$ (red dotted line) and therefore showing positivity is challenging when non-ideal conditions in the experiment are present. A possible solution to this would be to add two Markovian channels with all positive eigenvalues, as there may be room for deviation of $\lambda_{\rm min}$ from its ideal value while still remaining positive. Unfortunately, such an example is not known at present. Nevertheless, it is evident that the gap between the $\lambda_{\rm min}$ for the individual channels and the total channel increases with $t$, as expected from the theory. However, it is not possible to use this gap as a direct measure of the relative amount of non-Markovianity for the individual and joint channels.
\begin{figure*}[t]
\includegraphics[width=17cm]{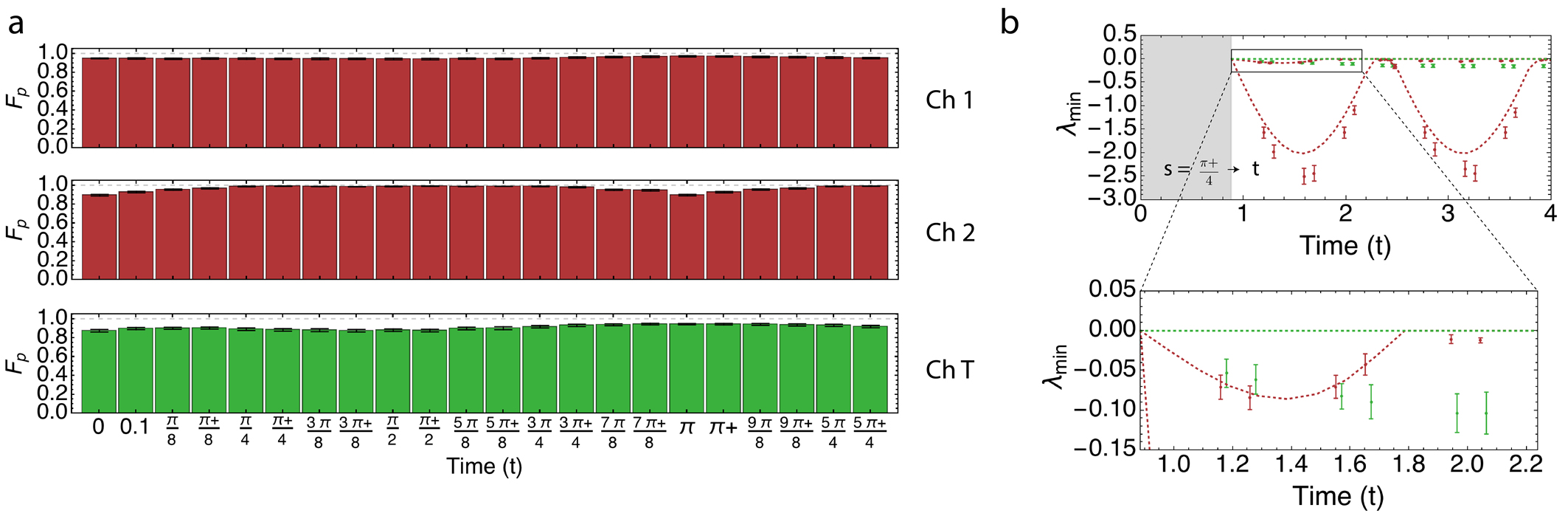}
\caption{Non-Markovian channel addition: ${\sf nM+nM=M}$. {\bf (a)} Process fidelities, $F_p$, for the $\chi$ matrices of the realized channels for the times corresponding to the dotted lines in Fig.~\ref{fig2}~(b). {\bf (b)} Minimum eigenvalues, $\lambda_{\rm min}$, of the Choi matrix for the intermediate maps from time $s$ to $t$, for $s=\pi^+/4=\pi/4+0.1$. The dashed grey area highlights that $t \geq s$ must be satisfied. The inset shows the finer details of $\lambda_{\rm min}$ for channel 1 and the total channel. The red (green) points correspond to the values from the experiment for the individual (total) channels. The red (green) dotted lines correspond to the ideal theoretical values for the individual (total) channels. The data points for channel 2 have been shifted slightly in time in the inset for clarity.}
\label{fig4} 
\end{figure*}

To measure the amount of non-Markovianity we use the measure introduced by Rivas, Huelga and Plenio (RHP)~\cite{Rivas10}, which has recently been given operational meaning~\cite{Anand19}. Here, the trace norm of the Choi matrix, $|| W(t+\epsilon,t)||_1$ (where $||A||_1={\rm Tr}\sqrt{A^\dag A}$), representing the map between two times $t$ and $t+\epsilon$ is used to form a measure, with $|| W(t+\epsilon,t)||_1=1$ iff $\Lambda_{t+\epsilon,t}$ is CP and $>1$ otherwise. This is equivalent to the positivity criteria introduced previously, as $|| W(t+\epsilon,t)||_1=\sum_k |\lambda_k|$, where the $\lambda_k$ are the eigenvalues of $W(t+\epsilon,t)$ and ${\rm Tr}(W(t+\epsilon,t))=\sum_k \lambda_k=1$. Thus, if the dynamics are Markovian then all the eigenvalues are positive and $|| W(t+\epsilon,t)||_1=1$ due to trace preservation of the underlying dynamics. Any deviation from this is non-Markovian and leads to $|| W(t+\epsilon,t)||_1>1$. While the two criteria of $\lambda_{\rm min}$ and the trace norm are equivalent, in the case of the latter the amount by which its value deviates from 1 can be used as a measure. Specifically, a time dependent function can be formed as $\bar{g}(t)={\rm Tanh}~g(t)$, where $g(t):= \lim_{\epsilon \to 0^+} \frac{||W(t+\epsilon,t)||_1-1}{\epsilon}$. A measure can then be defined over a time interval $I$ as ${\cal D}_{RHP}^I:\int_I \bar{g}(t) dt/\int_I \zeta \left[\bar{g}(t)\right] dt$, with $\zeta(x):=0$ if $x=0$ and 1 otherwise, together with the convention $0/0=0$~\cite{Rivas10}. The quantity ${\cal D}_{RHP}^I$ is then a measure of the non-Markovianity accumulated over the time interval $I$ in the periods when the dynamics was non-Markovian. 

In Fig.~\ref{fig3}~(c) we plot the function $\bar{g}(t)$ for the individual and total channels. The top panel shows the average values and error bar obtained from the set of 10,000 Choi matrices from the experiment for each pair of times $t$ and $t+0.1$, where we have used the approximation $\epsilon=0.1$. As before, the data points of channels 1 and 2 are shifted slightly off-center from each other for a given time $t$ for clarity. One can see that $\bar{g}(t)$ for the total channel roughly follows the expected theory behavior using $\epsilon=0.1$, as shown by the dotted green line. The difference between the theory value of $\bar{g}(t)$ for $\epsilon=0.1$ and $0^+$ is at most $2 \%$ over the range of $t$ considered. The data for the individual channels, however, do not follow the expected behavior of $\bar{g}(t)=0, \forall t$. Moreover, the integral of the area under a $\bar{g}(t)$ line is the measure ${\cal D}_{RHP}^I$, and assuming a similar behavior for points in between the ones measured, both the total channel and individual channels would show the accumulation of non-Markovianity during the dynamics.

When using the RHP measure, the non-ideal behavior of the individual channels could be the result of using $\epsilon=0.1$ as an approximation. Here, a problem arises for small $\epsilon$ values, as the channels in the experiment become similar to each other for small time differences. This similarity is made problematic by the sampling of the channels using quantum process tomography, as the qubits are encoded onto single photons whose count statistics are based on Poissonian fluctuations due to the SPDC source~\cite{James01}. This affects the error in the state tomography of the probe qubits and ultimately the spread of the 100 $\chi$ matrices obtained for a given channel, which results in the large error bars seen for $\bar{g}(t)$ in the top panel of Fig.~\ref{fig3}~(c). In principle it is possible to reduce these errors by increasing the measurement duration for state tomography to obtain a larger mean for the counts and thereby reduce the variance in the tomographic reconstruction. Unfortunately, it was not possible to observe an appreciable improvement within a reasonable amount of time in our current setup. 

\begin{figure*}[t]
\includegraphics[width=15cm]{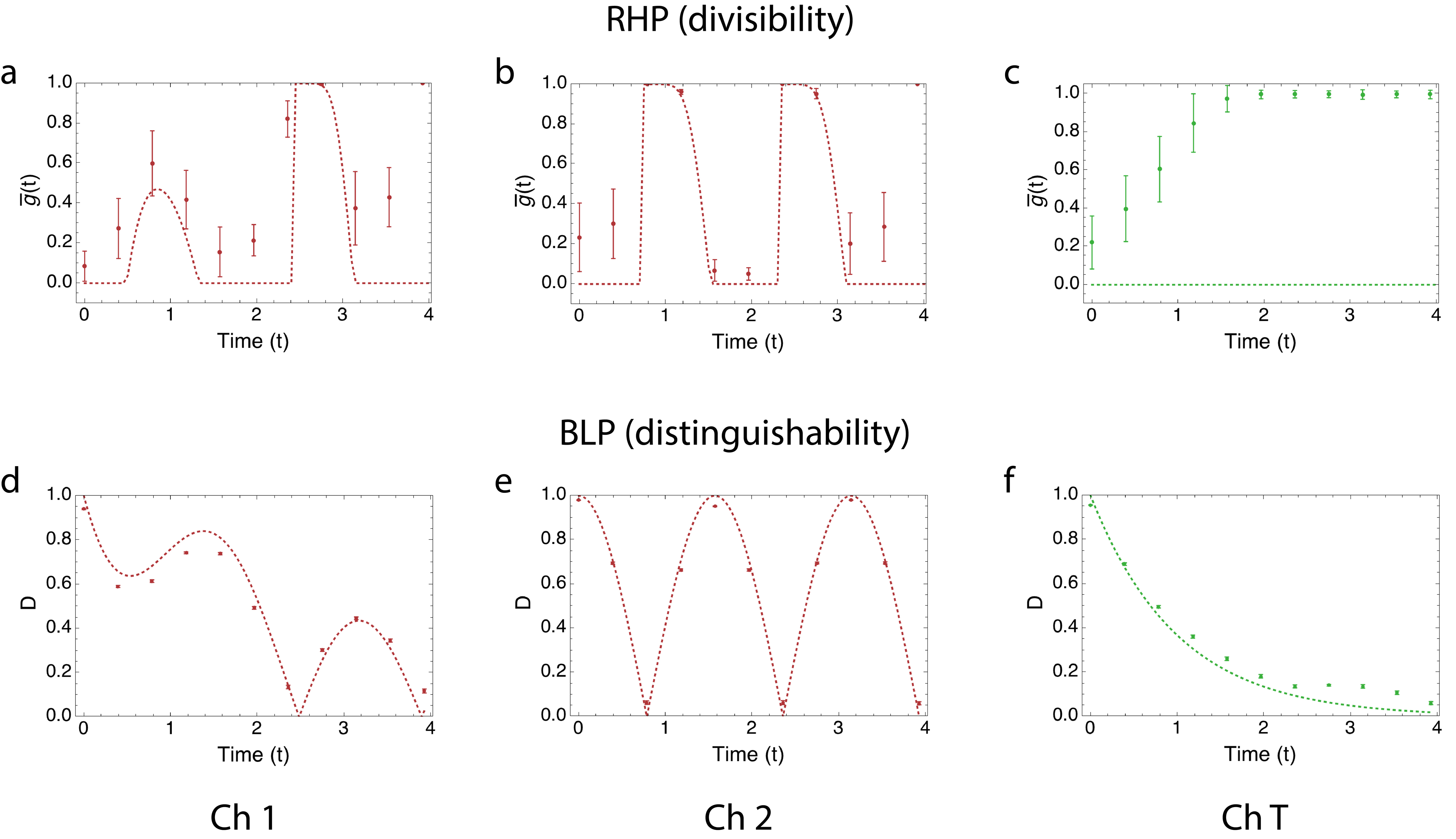}
\caption{Functions used for obtaining measures of non-Markovianity for ${\sf nM+nM=M}$. The upper row, (a)-(c), corresponds to the function $\bar{g}(t)$ used in the measure based on divisibility and the lower row, (d)-(f), corresponds to the trace distance $D(t)$ used in the measure based on distinguishability. The area below a given function corresponds to the amount of non-Markovianity accumulated during the dynamics. {\bf (a)} $\bar{g}(t)$ for channel 1. {\bf (b)} $\bar{g}(t)$ for channel 2. {\bf (c)} $\bar{g}(t)$ for the total channel. {\bf (d)} $D(t)$ for channel 1. {\bf (e)} $D(t)$ for channel 2. {\bf (f)} $D(t)$ for the total channel. The values and error bars are obtained by pairing up each of the 100 $\chi$ matrices from the experiment for time $t$ and the 100 $\chi$ matrices for $t+0.1$, giving $10,000$ Choi matrices for the intermediate map from $t$ to $t+0.1$. In all panels, the points correspond to the values from the experiment for the channels. The dotted lines correspond to the ideal theoretical values for the channels. The error bars in panels (d)-(f) are comparable to the points.}
\label{fig5} 
\end{figure*}

An alternative way of analysing $\bar{g}(t)$ with the aim of reducing the influence of the experimental fluctuations mentioned above is by taking the average $\chi$ matrix (of the 100 obtained) for each time from a pair for a given channel and calculating the Choi matrix for the intermediate map. This avoids having to calculate all 10,000 Choi matrices for the intermediate map, where the 100 $\chi$ matrices for each time can overlap appreciably with each other for small $\epsilon$ due to experimental statistical fluctuations. In the bottom panel of Fig.~\ref{fig3}~(c) we show the values of $\bar{g}(t)$ obtained using this method. Note that there are no error bars in the plot as only one value is obtained from a single pair of average $\chi$ matrices. Although not ideal, the trend of the data points is more inline with that expected from the theory. Thus, for the average channels realized in the experiment, the $\bar{g}(t)$ plots suggest that two `weak' non-Markovian channels (ideally Markovian) have been added to make a `stronger' non-Markovian channel. In order to make this statement more robust, more times in between those already taken would need to be acquired in order to obtain a smooth $\bar{g}(t)$ experimental line and enable ${\cal D}_{RHP}^I$ to be calculated accurately. By increasing the measurement duration for state tomography one may even be able to confirm the above addition for the non-averaged case using the 10,000 Choi matrices. 

%%%%%%%%%%%%%%%%%%%%%%%%%%%%
%%%%%%%%%%%%%%%%%%%%%%%%%%%%
\subsection{C. Non-Markovian channel addition}

In the second case, we follow the experimental procedure outlined in the previous sections in order to implement the addition of two non-Markovian channels to give a Markovian channel. For the fixed set of times marked in Fig.~\ref{fig2}~(b) we show the corresponding process fidelities in Fig.~\ref{fig4}~(a). As in the previous case, one can see that the experimental channels are of high quality for both the individual (Ch 1 and Ch 2) and total channels (Ch T), with the majority of process fidelity values $>0.90$. 

The lowest eigenvalue $\lambda_{\rm min}$ is plotted in Fig.~\ref{fig4}~(b) for time pairs starting with $s=\pi/4+0.1$ and $t>s$. This value of $s$ has been chosen as it leads to non-Markovian dynamics in both the individual channels, which is in contrast to the first case (${\sf M+M=nM}$), where any value of $s$ could be used to give non-Markovian dynamics in the total channel. One can see that for certain time pairs $\lambda_{\rm min}$ for the individual channels is negative, confirming that the channels are not divisible and therefore non-Markovian -- see inset of Fig.~\ref{fig4}~(b) for channel 1 in detail. The data points roughly follow the expected theory for the individual channels (red dotted lines). On the other hand, for the total channel, it can be seen in the inset that $\lambda_{\rm min}$ is still negative and the channel is therefore non-Markovian, even though it is expected to be Markovian in theory. As in the first case, the issue here is that ideally $\lambda_{\rm min}=0$ (green dotted line) and therefore showing positivity is challenging.

We next study the RHP measure of non-Markovianity for this case. In Fig.~\ref{fig5}~(a)-(c) we plot the function $\bar{g}(t)$ for the individual and total channels, where we have again used the approximation $\epsilon=0.1$. One can see that $\bar{g}(t)$ for the individual channels roughly follows the expected theory behavior (using $\epsilon=0.1$) as shown by the dotted red lines in Fig.~\ref{fig5}~(a) and (b).  Using $\epsilon=0.1$ instead of $0^+$ in the theory causes a small shift of $\bar{g}(t)$ in time of $\delta t \sim0.05$, which makes negligible difference to the expected theory behavior seen in the plot. On the other hand, the total channel does not follow the expected behavior, as shown by the dotted green line in Fig.~\ref{fig5}~(c). The integral of the area under a $\bar{g}(t)$ line is the measure ${\cal D}_{RHP}^I$, and assuming a similar behavior for points in between the ones measured, then both the individual and total channels would show the accumulation of non-Markovianity during the dynamics. A plot of $\bar{g}(t)$ based on the average $\chi$ matrix for each time of a pair for a given channel, as done for the first case, is not shown as the values correspond closely to those of the data points in Fig.~\ref{fig5}~(a)-(c). 

A possible reason for the increase in $\bar{g}(t)$ for the total channel as $t$ increases, is that for two different times $t$ and $t+\epsilon$, the $\chi$ matrices for the channel become similar to each other as $t$ increases and overlap considerably when experimental noise is included. This is because the total channel represents a phase damping channel along the $x$-axis and larger $t$ values give channels that becomes increasingly similar for a fixed $\epsilon$. As the values of $\bar{g}(t)$ for the total channel show strong non-Markovianity, it is unfortunately not possible to conclude that two non-Markovian channels have been added to make a Markovian channel, except from $t=0 \to 2$, where the total amount of non-Markovianity accumulated for the individual channels is larger than that for the total channel, with a total area ratio of approximately $1.5:1$. In this case, the plots suggest that the amount of non-Markovianity has been reduced by the addition of two non-Markovian channels. In other words, two `strong' non-Markovian channels have been added to make a `weaker' non-Markovian channel (ideally Markovian). As in the previous case, in order to make this statement more robust, more times in between those already taken would need to be acquired and the measurement duration for state tomography increased to reduce statistical fluctuations. Improvements to the experimental setup and the theoretical model are all good starting points for future work in this direction.

Interestingly, for this second case we are able to use a second criteria for witnessing non-Markovian dynamics which is based on the distinguishability measure introduced by Breuer, Laine and Piilo (BLP)~\cite{Breuer09} and uses the trace distance. Here, a single qubit at time $t$ can be written as $\rho(t)=\frac{1}{2}(\openone+\sum_{i=1}^3 x_i(t) \sigma_i)$. The trace distance between two states $\rho_1$ and $\rho_2$ evolved over a time duration $t$ is then $D(\rho_1,\rho_2)=\frac{1}{2}((x_{1,1}-x_{2,1})^2+(x_{1,2}-x_{2,2})^2+(x_{1,3}-x_{2,3})^2)^{1/2}$, where $x_{j,i}$ is the $i$-th Pauli component of the state $\rho_j$ evolved to time $t$. If the time derivative of the trace distance is positive for any time, then this is a witness of non-Markovian dynamics in terms of distinguishability. We emphasize that contrary to classical stochastic processes, there is no unique, universal definition of quantum Markovianity due to the absence of conditional probabilities. Thus, Markovianity can be defined in terms of divisibility or distinguishability, although these two definitions do not always coincide.

While non-Markovian dynamics in terms of divisibility is linked to the measurement process, the trace distance method based on distinguishability does not need to refer to measurements. In general, non-Markovian dynamics in terms of distinguishability imply non-Markovian dynamics in terms of divisibility but not vice-versa, as a map that is not divisible can result in a decreasing trace distance giving a negative time derivative. The trace distance therefore does not witness all non-Markovian dynamics. This is the reason the trace distance was not used in the first case, where Markovian channels were added, as the total non-Markovian channel is not expected to be theoretically witnessed. On the other hand, in the second case being studied here, while the trace distance may not help in determining the Markovian behavior of the total channel (in terms of divisibility), it is interesting to study its dependence for the individual and total channels, and compare it with the function $\bar{g}(t)$ used in the RHP measure. The trace distance is turned into a measure over the time period $I$ using the formula ${\cal D}_{BLP}^I={\rm max}_{\rho_1,\rho_2}\{ 0, \int_I \frac{d D(t)}{dt}dt \}$. 

In Fig.~\ref{fig5}~(d)-(f) we show the time dependence of $D(t)$ for the individual and total channels using the pair of states $\rho_1=\ketbra{H}{H}$ and $\rho_2=\ketbra{V}{V}$. Note that any pair of states can be used to provide a lower bound for ${\cal D}_{BLP}^I$. For the individual channels shown in Fig.~\ref{fig5}~(d) and (e) one can clearly see the behavior of $D(t)$ from the experiment matches the expected theory behavior (red dotted lines). Moreover, the behavior of $D(t)$ matches that of $\bar{g}(t)$: when $D(t)$ increases (positive time derivative) thereby witnessing non-Markovian dynamics in terms of distinguishability, the function $\bar{g}(t)$ measuring non-Markovian dynamics in terms of divisibility is positive also. 

On the other hand, when $D(t)$ decreases (negative time derivative) thereby witnessing Markovian dynamics in terms of distinguishability, the function $\bar{g}(t)$ is reduced close to zero corresponding to weak non-Markovian dynamics in terms of divisibility. For the total channel, one can clearly see the distinction between $D(t)$ in Fig.~\ref{fig5}~(f) which is decreasing and thereby witnessing Markovian dynamics in terms of distinguishability and $\bar{g}(t)$ in Fig.~\ref{fig5}~(c) which is far from zero corresponding to strong non-Markovian behavior in terms of divisibility. Thus, the trace distance appears to be a more robust witness of Markovianity in the experiment in terms of distinguishability. From the $D(t)$ plots for the individual and total channels in Fig.~\ref{fig5}~(d)-(f) we can confirm experimentally, for example between $t=\pi/4$ and $\pi/2$, that two non-Markovian channels have been added to make a Markovian channel, {\it i.e.} ${\sf nM}+{\sf nM}={\sf M}$, in terms of distinguishability. 

The experimental improvements outlined in the previous section for the first case may improve the situation for the total channel in this second case and allow ${\sf nM}+{\sf nM}={\sf M}$ to be verified in terms of divisibility, as already mentioned. It may also help to have an example where the total Markovian channel (when two non-Markovian channels are added) has all positive eigenvalues for its Choi matrices to allow for imperfections in the experiment. 

%%%%%%%%%%%%%%%%%%%%%%%%%%%%
%%%%%%%%%%%%%%%%%%%%%%%%%%%%
%%%%%%%%%%%%%%%%%%%%%%%%%%%%
%%%%%%%%%%%%%%%%%%%%%%%%%%%%
\section{IV. Summary} 

In this work we studied two cases for the addition of Markovian and non-Markovian channels in a photonic setup. The first case involved the addition of two Markovian channels and we studied the extent to which the individual channels were Markovian in the experiment and the total channel was non-Markovian. The second case involved the addition of two non-Markovian channels and we studied the extent to which the individual channels were non-Markovian in the experiment and the total channel was non-Markovian. In both cases, we found that using the negativity of the Choi matrix method to witness expected non-Markovian dynamics works well. However, when used to witness expected Markovian dynamics it did not work as well due to fluctuations in the data from the experimental setup and the matrices having a lowest eigenvalue of zero. 

We also studied two different methods for quantifying the amount of non-Markovianity in the channels. The first method based on divisibility was used in both cases and it was found that the method had varying success -- the best results were seen for the non-Markovian channels. The expected Markovian channels unfortunately displayed non-Markovian behavior. The second method based on distinguishability was used in the second case only and worked well, confirming that the individual channels were non-Markovian in terms of distinguishability and divisibility and the total channel was Markovian in terms of distinguishability. It was not possible to confirm the total channel was Markovian in terms of divisibility, except that there was a smaller amount of non-Markovianity accumulated over a given period of time compared to the total amount from the individual channels.

Future work on the experimental addition of Markovian and non-Markovian channels would certainly benefit from improving the experimental setup. Furthermore, finding examples where the Markovian channels have all positive eigenvalues for the intermediate Choi maps would be beneficial. These factors would potentially allow a confirmation of the Markovian behavior in terms of divisibility for both cases and thus provide a firm demonstration of the non-convexity of the set of Markovian and non-Markovian channels. These experimental results confirm that Markovianity is an extremely fragile property (as reported in~\cite{Wudarski17}). Therefore, one needs to be careful while realizing/simulating Markovian behavior, because small discrepancies can lead to non-Markovian characteristics. 

%%%%%%%%%%%%%%%%%%%%%%%%%%%%
%%%%%%%%%%%%%%%%%%%%%%%%%%%%
%%%%%%%%%%%%%%%%%%%%%%%%%%%%
%%%%%%%%%%%%%%%%%%%%%%%%%%%%
{\it Acknowledgments.---} We thank T. Tashima for discussions during the early stages of the work. This research was supported by the South African National Research Foundation, the National Laser Centre, the UKZN Nanotechnology Platform, the South African National Institute for Theoretical Physics, and the South African Research Chair Initiative of the Department of Science and Innovation and National Research Foundation.

%%%%%%%%%%%%%%%%%%%%%%%%%%%%
%%%%%%%%%%%%%%%%%%%%%%%%%%%%
%%%%%%%%%%%%%%%%%%%%%%%%%%%%
%%%%%%%%%%%%%%%%%%%%%%%%%%%%

\end{document}